\documentclass{INTERSPEECH2023}
\usepackage{array}
\newcommand{\PreserveBackslash}[1]{\let\temp=\\#1\let\\=\temp}
\newcolumntype{C}[1]{>{\PreserveBackslash\centering}p{#1}}
\newcolumntype{R}[1]{>{\PreserveBackslash\raggedleft}p{#1}}
\newcolumntype{L}[1]{>{\PreserveBackslash\raggedright}p{#1}}

\interspeechcameraready

\title{ACA-Net: Towards Lightweight Speaker Verification using Asymmetric Cross Attention}
\name{Jia Qi Yip$^1$$^2$, Tuan Truong$^2$, Dianwen Ng$^1$$^2$, Chong Zhang$^1$, Yukun Ma$^1$, Trung Hieu Nguyen$^1$, Chongjia Ni$^1$, Shengkui Zhao$^1$, Eng Siong Chng$^2$, Bin Ma$^1$} 
\address{
  $^1$Alibaba Group\\
  $^2$Nanyang Technological University, Singapore\\
  }
\email{jiaqi006@e.ntu.edu.sg}

\begin{document}

\maketitle
 
\begin{abstract}
In this paper, we propose ACA-Net, a lightweight, global context-aware speaker embedding extractor for Speaker Verification (SV) that improves upon existing work by using Asymmetric Cross Attention (ACA) to replace temporal pooling. ACA is able to distill large, variable-length sequences into small, fixed-sized latents by attending a small query to large key and value matrices. In ACA-Net, we build a Multi-Layer Aggregation (MLA) block using ACA to generate fixed-sized identity vectors from variable-length inputs. Through global attention, ACA-Net acts as an efficient global feature extractor that adapts to temporal variability unlike existing SV models that apply a fixed function for pooling over the temporal dimension which may obscure information about the signal's non-stationary temporal variability. Our experiments on the WSJ0-1talker show ACA-Net outperforms a strong baseline by 5\% relative improvement in EER using only 1/5 of the parameters.
\end{abstract}
\noindent\textbf{Index Terms}: Speaker Verification, Asymmetric Cross Attention, Lightweight

\section{Introduction}
Speaker Verification (SV) is the task of determining if a given speech segment belongs to a claimed enrolled speaker. This task is typically achieved by the comparison of fixed-length speaker embeddings computed from variable-length utterances. For speaker verification, a speaker embedding extractor must produce close embeddings for different utterances of the same speaker, and distant embeddings for utterances from different speakers. In addition, speaker embeddings can also be used in other speech domains such as speaker diarization~\cite{ECAPA_diarization} and speaker extraction~\cite{chenglin} to create a speaker targeted frontend to ASR models like~\cite{dehubert, chen2023leveraging} or for keyword spotting~\cite{cnvmix}.

Temporal statistics pooling~\cite{x-vectors, statpool, deepsets} is commonly used by embedding extraction networks in SV models to handle variable input lengths. Temporal statistics pooling refers to the channel-wise pooling of a model's embedding vector, commonly by taking the mean, max or standard deviation of all time steps in that channel, to obtain a single representative value for that channel. However, depending on the statistics used, the pooling method may obscure variability across time steps that may be important in discriminating between speakers. Additionally, statistics pooling assumes that the speech signal has statistical properties that remain stationary over time, which may not always hold true. Recent models such as RawNet3~\cite{RawNetv3}, LargeResNet-MagFace~\cite{Kuzmin_2022}, MFA-Conformer~\cite{MFA-Conformer} and ECAPA-TDNN~\cite{ECAPA-TDNN} have used context-aware or attentive statistics pooling~\cite{attn_stat_pool} to vary the weight of each time-step during the pooling operation.

\begin{figure}[t]
  \centering
  \includegraphics[width=\linewidth]{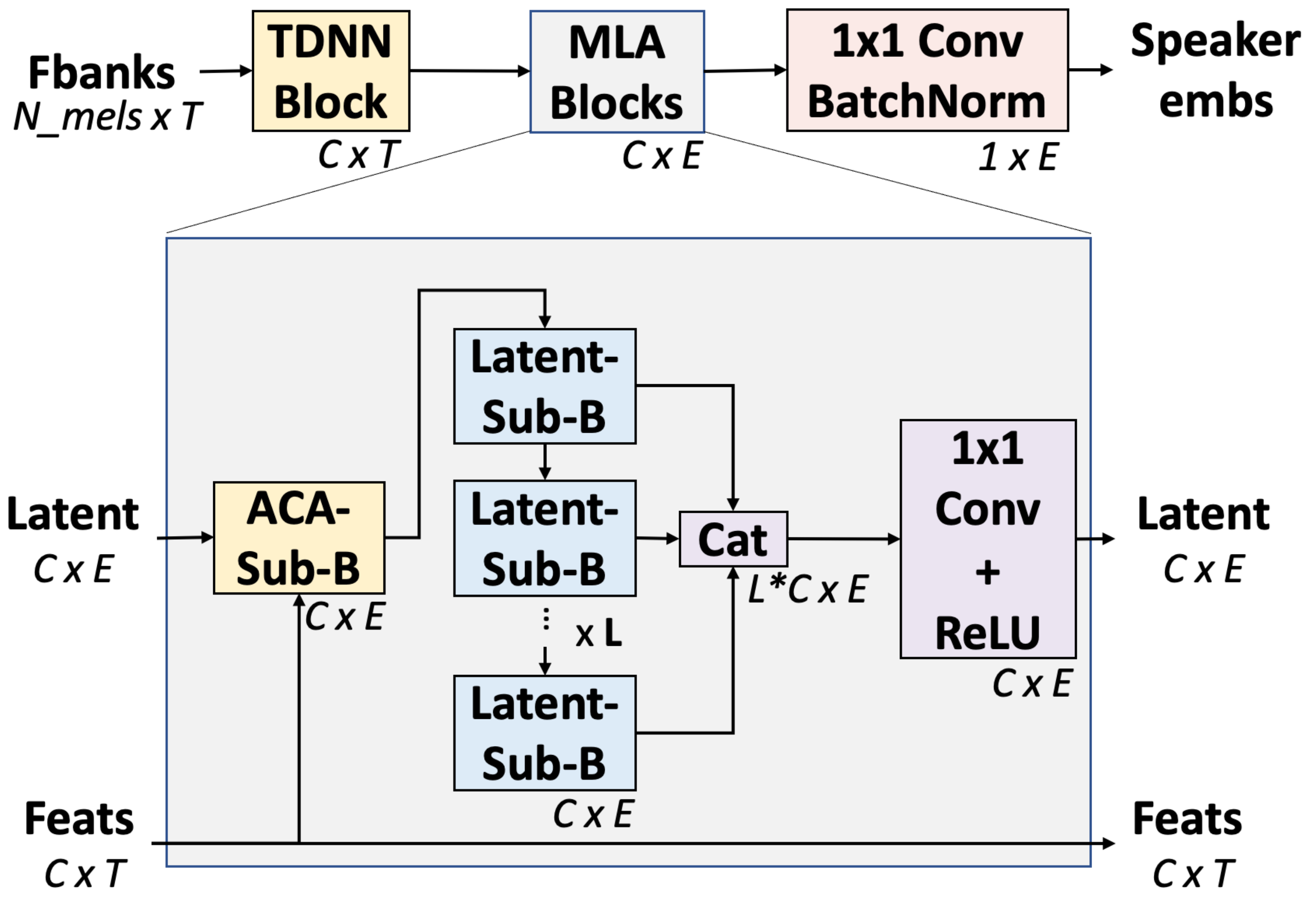}
  \caption{Overall architecture of ACA-Net. The model consists of a single 1x1 TDNN block, followed by the Multi-Layer Aggregation (MLA) block. The details of the MLA block are shown in the bottom part of the figure. The MLA block accepts a Latent and Features (Feats) as inputs and outputs a new Latent vector. The Features remain unchanged by the block. The ACA- and Latent-Sub-Blocks (Sub-B) are detailed in Figure~\ref{fig:cross-and-latent}.}
  \label{fig:ACA-Net-overview}
  \vspace{-15pt}
\end{figure}

Besides temporal pooling, it has been shown that embedding extractors benefit from having global information even while modeling the local features in each time step~\cite{MFA-Conformer}. For fully convolutional models, the Squeeze-and-Excitation layer~\cite{SEBlock} in the commonly used Res2Net block~\cite{Res2Net} was used to bring global information into the model. More recently, the transformer architecture was used to encode global information in MFA-Conformer~\cite{MFA-Conformer}. Elsewhere, transformers~\cite{TRANSFORMER} have also been applied successfully in many speech applications such as speech separation~\cite{subakan2021sepformer}, and speech enhancement~\cite{SepformerEnhancement}. Nevertheless, a drawback of using transformers is the high cost of computing self-attention over large matrices.

\begin{figure}[t]
  \centering
  \includegraphics[width=\linewidth]{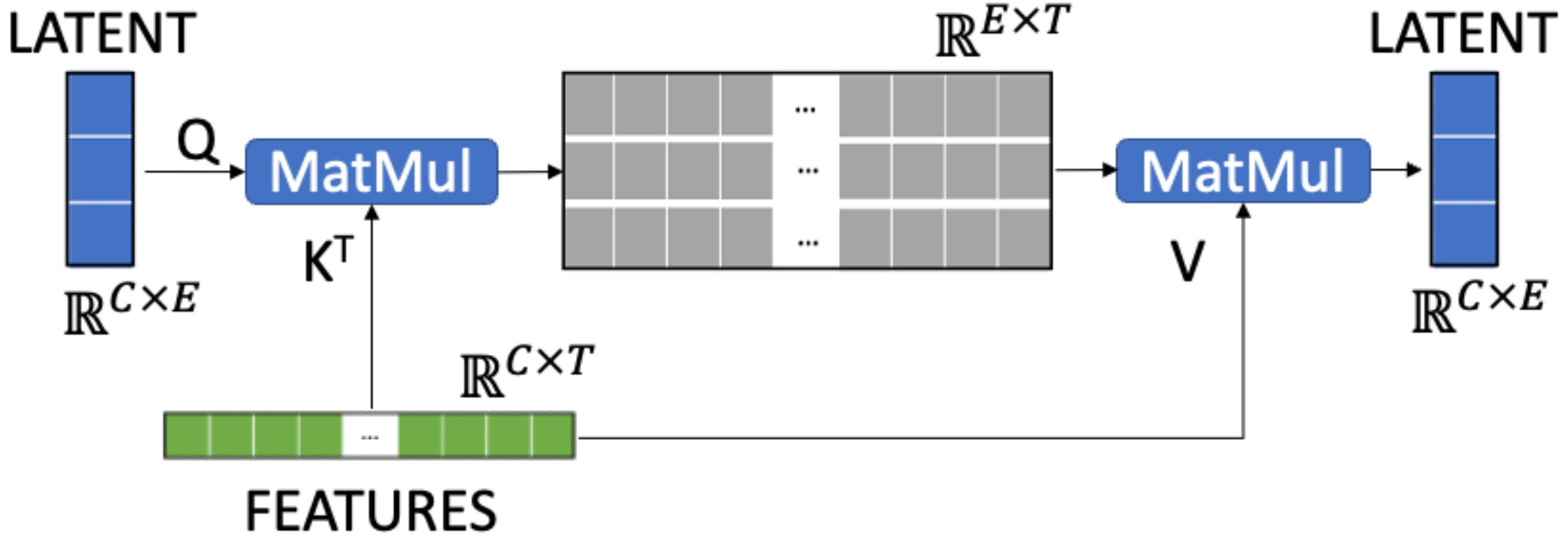}
  \caption{Illustration of Asymmetric Cross Attention (ACA). Applying the $(QK^{T})V$ series of matrix multiplications (MatMul) in the attention operation using a fixed-sized Query (Q) with variable length Key (K) and Values (V) always results in an output of the same dimensions as Q.}
  \label{fig:ACA-diagram}
  \vspace{-17pt}
\end{figure}

Here we propose ACA-Net, as shown in Figure~\ref{fig:ACA-Net-overview}, which uses Asymmetric Cross Attention (ACA)~\cite{setAttention, Perceiver} to avoid the high computational cost of self-attention while eliminating the need for temporal pooling. ACA computes attention between a small latent query and a large feature sequence as the key and value matrices, shown in Figure~\ref{fig:ACA-diagram}. This distils the temporal dimension of the feature input down to the embedding dimension. ACA-Net is thus a lightweight, computationally efficient model with strong global context modeling which can capture more fine-grained information about the speech signal~\cite{subakan2021sepformer} and provides better discrimination between speakers~\cite{MFA-Conformer}. The obtained latent is then refined through Multi-Layer Aggregation (MLA) over multiple self-attention sub-blocks, further enhancing speaker verification performance. Experiments on WSJ0-1talker~\cite{chenglin} show that the proposed ACA-Net surpasses strong baselines such as ECAPA-TDNN~\cite{ECAPA-TDNN} and RawNet3~\cite{RawNetv3} while using only 1/5 of the parameters.

While ACA-based methods have recently been used in a variety of domains, including natural language processing~\cite{ACANLP, ACANLP2}, remote sensing~\cite{ACAHNet}, image-text matching~\cite{ACAGAAN} and most notably in the Perceiver class of neural networks~\cite{Perceiver, PerceiverIO, PerceiverAR, HiP}, to our knowledge ACA-Net is the first speaker embedding extractor to use ACA instead of the temporal pooling methods commonly used in current SV models.


\vspace{-2pt}
\section{Methodology}

\begin{figure*}[th!]
    \centering
    \includegraphics[width=480pt]{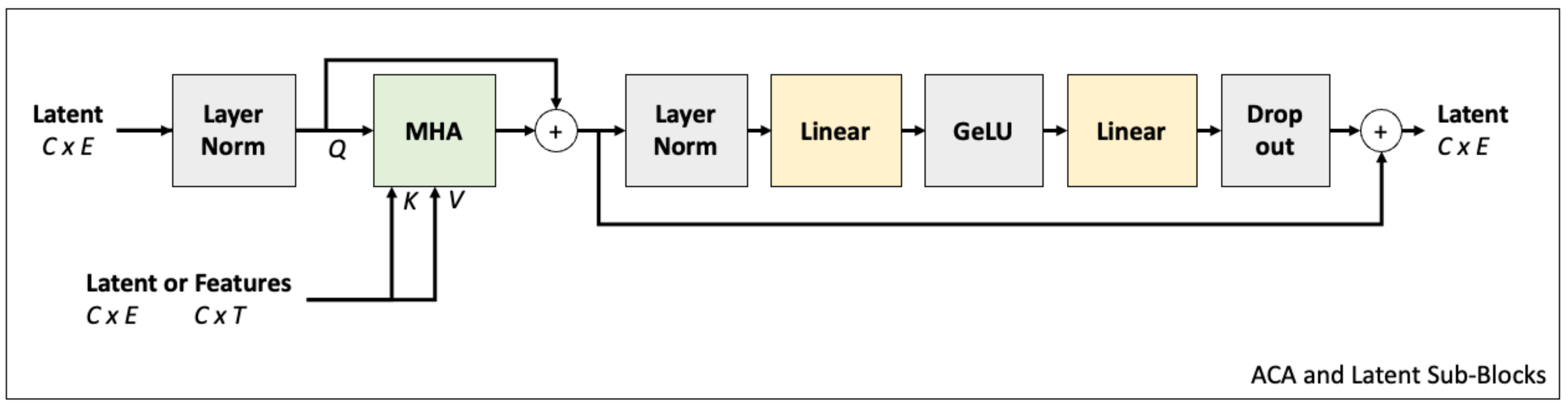} 
    \caption{Detailed view of the ACA- and Latent-Sub-Blocks within an MLA block represented as``ACA-Sub-B" and ``Latent-Sub-B" in the overall architecture of ACA-Net shown in Figure~\ref{fig:ACA-Net-overview}. ACA- and Latent-Sub-Blocks share the same architecture but differ in their inputs to the Multi-head Attention (MHA) layer. The ACA sub-block passes a Latent into the Query (Q) and Features into the Key (K) and Value (V) of the MHA while the  Latent-Sub-Block performs self-attention on the same Latent vector.}
    \label{fig:cross-and-latent}
    \vspace{-10pt}

\end{figure*}

The overall architecture of the proposed ACA-Net\footnotemark{} is shown in Figure~\ref{fig:ACA-Net-overview}. The model accepts audio input processed through a filterbank and consists of a single TDNN block followed by the Multi-Layer Aggregation (MLA) Block and a final 1x1 convolution used to reduce the channel dimension back to 1 for the final embedding.
\footnotetext{The model is available at \href{https://github.com/Yip-Jia-Qi/ACA-Net}{github.com/Yip-Jia-Qi/ACA-Net}}

\vspace{-2pt}
\subsection{TDNN Block}
The TDNN block used after the filterbank layer follows the implementation of~\cite{ECAPA-TDNN} in~\cite{speechbrain}. The TDNN block in ACA-Net consists of a single depth-wise 1D Convolutional layer, followed by ReLU activation and 1D batch normalization. The purpose of this block is to serve as a further feature extractor to decouple the number of filterbanks from the input channels to the MLA block.

\vspace{-3pt}
\subsection{Asymmetric Cross Attention}
The ACA sub-block shown in Figure~\ref{fig:cross-and-latent} makes use of the standard Multi-Head Attention (MHA) as per Pytorch, which can be defined as follows:
\begin{equation}
\begin{aligned}
   \text{MultiHead}(Q,K,V) &= \text{Concat}(head_{1}, \dots head_{h})W^{O} \\
   head_{i} &= \text{Attention}(QW^{Q}_{i},KW^{K}_{i},VW^{V}_{i}) \\
   \text{Attention}(Q,K,V) &= \text{softmax}(\frac{QK^{T}}{\sqrt{d}})V
   \end{aligned}
\end{equation} 

\noindent where $d$ denotes the number of channels and $Q$, $K$, $V$ denote the Query, Key and Value of the MHA respectively. $W^{Q}_{i}$, $W^{K}_{i}$, $W^{V}_{i}$,$W^{O}_{i}$ are the projection parameter matrices.

The standard transformer makes use of MHA where $Q \in \mathbb{R}^{C\times T}$, $K \in \mathbb{R}^{C\times T}$ and $V \in \mathbb{R}^{C\times T}$, whereas ACA uses a $Q \in \mathbb{R}^{C\times E}$ where $E\lll T$. Furthermore, E is the embedding size which is a fixed hyperparameter while T is the time dimension which is variable. C represents the number of channels. While the output of MHSA will be $\mathbb{R}^{C\times T}$ the output of ACA will be $\mathbb{R}^{C\times E}$, which is a much smaller latent vector. Importantly, the dimensions of the ACA output will be independent of $T$.

\subsection{ACA and Latent Sub-Blocks}
As shown in Figure~\ref{fig:cross-and-latent}, the ACA sub-block and the latent sub-block share the same block architecture. The key difference between the sub-blocks is the dimensions of the $K$ and $V$ inputs. The $Q$, $K$ and $V$ for the latent sub-block are the latent produced by the ACA sub-block. The $K$ and $V$ for the ACA sub-block is the feature sequence while $Q$ comes from random initialization. Additionally, for the ACA sub-block, sinusoidal positional encoding is added to the feature sequence before being passed into the MHA layer.

When given Feature input of dimensions $\mathbb{R}^{C\times T}$ the ACA sub-block reduces the time dimension to the embedding size resulting in an output of $\mathbb{R}^{C\times E}$. Meanwhile, since the latent sub-block performs self-attention on the latent, it results in no change in dimensions to the latent of $\mathbb{R}^{C\times E}$.

\subsection{The MLA Block}
The MLA block as shown at the bottom of Figure~\ref{fig:ACA-Net-overview} represents our key contribution. It consists of a single ACA sub-block (ACA-Sub-B) followed by a variable number of latent sub-blocks. Both blocks share the same design shown in Figure~\ref{fig:cross-and-latent}. Finally, the concatenated outputs of the latent sub-blocks are passed into a depth-wise 1D convolution and batch normalization layer. 
\begin{equation}
\begin{aligned}
   Latent_{init} & \sim \mathcal{N}(\mu, \sigma^{2}) \\
   Latent^{'} &= \text{ACA-Sub-B}(Latent_{init},Features)\\
   Latent^{''} &= \text{MLA}(Latent^{'}) \\
   Latent_{out} &= \text{ReLU}(\text{Conv1D}(Latent^{''})) \\
   \end{aligned}
\end{equation} 
\noindent where $\mathcal{N}$ denotes a truncated normal distribution with mean of $\mu$ and standard deviation of $\sigma^{2}$. $Latent \in \mathbb{R}^{C\times E}$ is randomly initialised and $Features \in \mathbb{R}^{C\times T}$ refers to the output from the TDNN block.

The purpose of the ACA sub-block is to compute an initial latent while the latent sub-block (Latent-Sub-B) refines the latent. The latent is refined through MLA, where the latent vector is passed through multiple latent sub-blocks, with the output of each latent sub-block aggregated by concatenation along the channel dimension and passed through a depth-wise convolution at the end to return the channel dimension back to its original size. MLA can be described as follows:
\begin{equation}
\begin{aligned}
    &\text{MLA}(L) = \text{Conv1D}(\text{Concat}(Layer_{1}, \dots Layer_{j})) \\
    &\text{for } Layer_{j+1} = \text{Latent-Sub-B}(Layer_{j}) \\
   \end{aligned}
\end{equation} 

\noindent where $L = Layer_{0}\in \mathbb{R}^{C\times E}$ and $j$ denotes the number of latent sub-blocks.

This MLA is similar to the multi-scale feature aggregation method employed in~\cite{MFA-Conformer} and~\cite{ECAPA-TDNN} although the latents are not multi-scale since the ACA produces the latent using the full global context, resulting in only a single scale.

\section{Experiments}
\subsection{Dataset}

For all experiments, we train and evaluate the models on the WSJ0-1talker speaker verification dataset~\cite{chenglin} which is drawn from the WSJ0 corpus~\cite{wsj0}. This dataset is designed to test speaker verification performance for speaker embedding extractors to be used in speaker extraction systems such as~\cite{spex}, where a lightweight speaker embedding extractor like ACA-Net is important since the speaker embedding is auxiliary to the main speech separation network. Speaker extraction is commonly benchmarked on the WSJ0-2mix~\cite{wsj0-2mix} dataset, which consists of mixtures generated from the WSJ0 Corpus~\cite{wsj0}. 

The WSJ0-1talker dataset~\cite{chenglin} consists of 101 speaker training and development sets drawn from the ``si\_tr\_s" collection from the WSJ0 corpus~\cite{wsj0} and a test set of 18 speakers drawn from the``si\_dt\_05" and ``si\_et\_05" collections. While the training (20,000 utterances) and development (5,000 utterances) sets share speakers but have different utterances, the testing (3,000 utterances) set consists of 18 separate speakers unseen during training. The verification pairs for testing are randomly selected from the training set. All utterances were down-sampled from 48kHz to 8kHz.

\subsection{Experimental Setup}
All models were trained for 25 epochs using the Adam optimizer~\cite{adamoptim} with a Cyclical Learning Rate Scheduler~\cite{CLRS} with a base learning rate of $10^{-7}$ and a maximum learning rate of $10^{-2}$. For larger models, we use a maximum learning rate of $10^{-3}$ for better stability. The batch size was set to 32 or reduced to 16 on larger models due to memory limitations. All experiments were done on 1 GPU with 16GB RAM.

The loss function used for all model training was Additive Angular Margin (AAM), or ArcFace~\cite{aamsoftmax} with a margin of 0.2 and a scale of 30. During Testing, use the Equal Error Rate (EER) metric and the minimum detection cost function (minDCF) metric to measure performance. The minDCF metric is calculated with the hyperparameters $P_{target}$=0.01 and $C_{falsealarm}$ = $C_{miss}$=1. AAM, EER and minDCF functions are implemented by the Speechbrain~\cite{speechbrain} training framework.

\subsection{Model Hyperparameters}
We train two baseline models, ECAPA-TDNN~\cite{ECAPA-TDNN} and RawNet3~\cite{RawNetv3}, which have not previously been reported on the WSJ0-1talker dataset, in addition to ACA-Net. All models have been trained on the Speechbrain~\cite{speechbrain} framework.

\vspace{3pt}
\noindent \textbf{ECAPA-TDNN.} The ECAPA-TDNN model~\cite{ECAPA-TDNN} is a recent state-of-the-art model incorporating time-delay neural networks (TDNN) and Multi-scale Feature Aggregation across three layers. We use the existing implementation of the ECAPA-TDNN model included as part of the Speechbrain~\cite{speechbrain} framework. All model hyperparameters are set per the defaults in Speechbrain.

\vspace{3pt}
\noindent \textbf{RawNet3.} To train the RawNet3 model we made minimal modifications to the code provided by the authors of~\cite{RawNetv3} for it to work in Speechbrain. We train two versions of the models by adjusting the ``C" hyperparameter which controls the number of channels in the convolutional layers of the model. The original model in~\cite{RawNetv3} has C=1024 while we train a smaller version with C=512 for a model that is closer in parameter size to ACA-Net.

\vspace{3pt}
\noindent \textbf{ACA-Net.} The base ACA-Net model consists of 1 MLA block with 1 ACA sub-block and 3 latent sub-blocks. The embedding size of the base ACA-Net model is 512. Throughout all experiments, the dropout of the sub-blocks is set to 0.2, the channel dimension is fixed at 256 with the size of all linear layers is set to 1024. Input features are derived using a filter bank with 80 filters, a hop length of 10 and a window length of 25. Positional Encoding is added to the features using a standard sinusoidal positional encoding function. Where applicable, the initial latent for ACA layers within the ACA sub-blocks were initialized according to a truncated normal distribution with mean 0, standard deviation 0.02, and truncation bounds [-2, 2]

\begin{table}[h!]
\vspace{-5pt}
  \caption{Performance comparison of ACA-Net against other popular SV models on the WSJ0-1talker verification test set. Number of parameters for the model are reported where available. SV-T and RawNet3 are time domain models while the rest operate in the frequency domain.\vspace{-5pt}}
  \label{tab:baselines}
  \centering
  \begin{tabular}{p{3.1cm}C{0.9cm}C{0.6cm}C{1.1cm}}
    \toprule
    \textbf{Model} & \textbf{Params (M)} & \textbf{EER$\downarrow$ (\%)} & \textbf{minDCF$\downarrow$}\\
    \midrule
    x-vector~\cite{xvecPLDA}& - & 5.87 & 0.69 \\
    SV-T~\cite{chenglin} & - & 4.40 & 0.45 \\
    SV-F~\cite{chenglin} & - & 4.37 & 0.42 \\
    RawNet3 (C=512)~\cite{RawNetv3}   &  6.5 & 3.94 & 0.38 \\
    RawNet3 (C=1024)~\cite{RawNetv3}   &  16.3 & 3.46 & 0.38 \\
    ECAPA-TDNN~\cite{ECAPA-TDNN} & 20.8 & 2.99 & 0.32  \\
    SV-FA~\cite{chenglin} & - & 2.90 & 0.36 \\
    \midrule
    ACA-Net   &  3.6 & \textbf{2.85} & \textbf{0.31} \\
    \bottomrule
  \end{tabular}
  \vspace{-10pt}
\end{table}

\subsection{Experimental Results}
In Table~\ref{tab:baselines} we report the verification performance of ACA-Net on the WSJ0-1talker dataset compared with two reimplemented baselines ECAPA-TDNN and RawNet3. Additionally, we also compare the results against existing baselines, x-vector PLDA, SV-T, SV-F, and SV-FA reported in~\cite{chenglin}, although the number of parameters for the existing baselines was not reported.

Based on the WSJ0-1talker dataset as shown in Table~\ref{tab:baselines}, ACA-Net achieves the lowest EER and minDCF out of all the models while using only 1/5 of the parameters of ECAPA-TDNN and RawNet3. Except for ACA-Net, all baseline models make use of some sort of temporal pooling method. The x-vector model, an older model from 2018~\cite{x-vectors}, unsurprisingly performs the worst without the benefit of recent innovations. When comparing the results of SV-T with SV-F as well as the results of ECAPA-TDNN and SV-FA with RawNet3, we find that the frequency domain approaches have an advantage over time domain models. This aligns with findings in~\cite{RawNetv3}. Additionally, the result that the 512-channel version of RawNet3 with fewer parameters performs more poorly than the 1024-channel version suggests that the poorer performance of the larger baseline models, ECAPA-TDNN and RawNet3, is not simply due to the large models overfitting on the WSJ0-1talker dataset.

\subsection{Ablation Study}
We conduct an ablation study on various components of the design of ACA-Net to show their contribution to the performance of the model in Table~\ref{tab:design_ablations}. In the subtractive ablation experiments, we remove the concatenation step of the MLA block, leaving the output of the last latent sub-block to be passed through a depth-wise convolution with no change in dimension. We also experimented with removing the positional encoding of the features before the ACA sub-block as well as removing latent sub-block from the MLA block. In the additive ablation experiments, we used weight sharing across the 3 latent sub-blocks.

The drop in performance after the removal of MLA and latent sub-blocks validates the importance of these design features in the model. Since the model relies on attention, performance falls when the positional encoding of the features is removed because information about the relative positions of the input features is lost. The relative position of input features is important because speaker identity is determined through speech patterns that are only identifiable if they occur in sequence. Weight sharing across the latent sub-blocks, turns the latent refinement into a recursive processes, resulting in a 40\% reduction in parameter size. While this is accompanied by a drop in performance, we note that this smaller model still achieves EER on par with the 1024-channel RawNet3 model from Table\ref{tab:baselines}.

\begin{table}[htbp]
  \caption{Ablation study of ACA-Net with the decomposition of different components in terms of parameter sizes, EER and minDCF, respectively.}
  \label{tab:design_ablations}
  \centering
  \begin{tabular}{p{3.1cm}C{0.9cm}C{0.6cm}C{1.1cm}}
    \toprule
    \textbf{Model} & \textbf{Param (M)} & \textbf{EER$\downarrow$ (\%)}  & \textbf{minDCF$\downarrow$} \\
    \midrule
    ACA-Net & 3.6 & 2.85 & 0.31 \\
    \midrule
    - MLA & 3.3 & 3.90 & 0.24 \\
    - Latent-Sub-Blocks   &  3.6 & 4.68  & 0.42 \\
    - Positional Encoding   &  3.6 & 4.68 & 0.47 \\
    \midrule
    + weight sharing & 2.0  & 3.46  & 0.45    \\
    \bottomrule
  \end{tabular}
\end{table}

\begin{figure}[ht!]
  \centering
  \includegraphics[width=180pt]{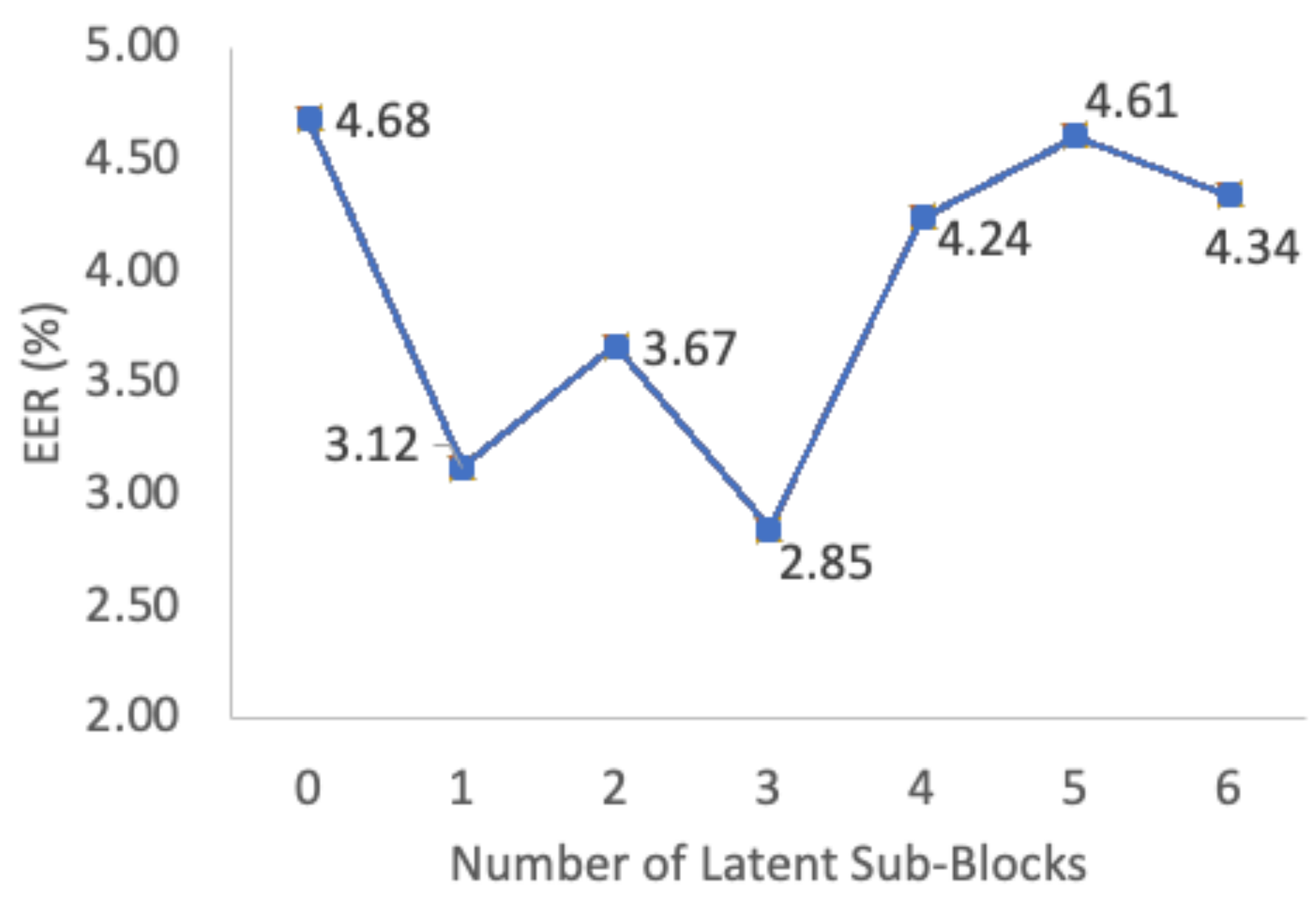}
  \caption{Ablation Study over a number of latent sub-blocks in the base ACA-Net. Increasing the number of blocks increases the depth of the model.}
  \label{fig:latent-attention-ablation}
  \vspace{-15pt}
\end{figure}

\vspace{5pt}
Next, we conduct a series of ablation experiments on various model dimensions of the base ACA-Net. Specifically, we experiment with changing the number of latent sub-blocks (Figure~\ref{fig:latent-attention-ablation}) to determine if more latent sub-blocks can result in better performance and embedding size (Table~\ref{tab:embedding_ablations}) to determine if giving the model more space to place speakers can help it better differentiate speakers.

Based on the results shown in Figure~\ref{fig:latent-attention-ablation}, the number of latent sub-blocks for ACA-Net seems to be optimal at 3 sub-blocks, with significantly worse performance observed beyond 3 sub-blocks. One possibility is that after the 3rd sub-block, the additional parameters introduced do not contribute to the further refinement of the latent as there are now a too many layers between these deeper sub-blocks and the original features distilled by the ACA sub-block.

\begin{table}[h!]
    \vspace{-5pt}
  \caption{Verification performance of ACA-Net with different embedding sizes (i.e., $E=256, 512, 1024$). Embedding size determines the length of the vector used by the model to discriminate between speakers during speaker verification.}
  \label{tab:embedding_ablations}
  \centering
  \begin{tabular}{p{3.1cm}C{1.2cm}C{1.2cm}}
    \toprule
    \textbf{Model} & \textbf{EER$\downarrow$ (\%)}  & \textbf{minDCF$\downarrow$} \\
    \midrule
    ACA-Net (E=256)  & 5.16 & 0.32 \\
    ACA-Net (E=512) & 2.85 & 0.31 \\
    ACA-Net (E=1024) & 4.96 & 0.49 \\
    \bottomrule
  \end{tabular}
\vspace{-5pt}
\end{table}

In SV, is has been shown that embedding dimension size is an important hyperparameter since it determines the volume of high-dimensional space available in the vector used to discriminate between speakers~\cite{Gu_2021}. Increasing embedding vector size could improve discriminability by increasing the volume of space available for the model to encode speakers, however, having a space that is too large could also result in utterances from the same speaker being wrongly separated~\cite{Gu_2021}. We see this effect in Table~\ref{tab:embedding_ablations} where embedding sizes larger and smaller than the 512 used by the base ACA-Net result in worse performance.


\vspace{-5pt}
\section{Conclusion}
Here we presented ACA-Net, a lightweight speaker embedding extractor for SV. ACA-Net is the first model to apply ACA to SV and achieves impressive performance by replacing temporal pooling with global feature extraction through attention. On the WSJ0-1talker dataset, ACA-Net outperforms strong baselines, ECAPA-TDNN and RawNet3, on both EER and minDCF, despite using only 1/5 of the parameters. Overall, our experiments highlight the potential of ACA as an alternative to typical temporal pooling methods. 

\vspace{-5pt}
\section{Acknowledgements}
This work was supported by Alibaba Group through Alibaba Innovative Research (AIR) Program and Alibaba-NTU Singapore Joint Research Institute (JRI), Nanyang Technological University, Singapore.

\bibliographystyle{IEEEtran}
\bibliography{mybib}

\begin{thebibliography}{10}
\providecommand{\url}[1]{#1}
\csname url@samestyle\endcsname
\providecommand{\newblock}{\relax}
\providecommand{\bibinfo}[2]{#2}
\providecommand{\BIBentrySTDinterwordspacing}{\spaceskip=0pt\relax}
\providecommand{\BIBentryALTinterwordstretchfactor}{4}
\providecommand{\BIBentryALTinterwordspacing}{\spaceskip=\fontdimen2\font plus
\BIBentryALTinterwordstretchfactor\fontdimen3\font minus
  \fontdimen4\font\relax}
\providecommand{\BIBforeignlanguage}[2]{{%
\expandafter\ifx\csname l@#1\endcsname\relax
\typeout{** WARNING: IEEEtran.bst: No hyphenation pattern has been}%
\typeout{** loaded for the language `#1'. Using the pattern for}%
\typeout{** the default language instead.}%
\else
\language=\csname l@#1\endcsname
\fi
#2}}
\providecommand{\BIBdecl}{\relax}
\BIBdecl

\bibitem{ECAPA_diarization}
N.~Dawalatabad, M.~Ravanelli, F.~Grondin, J.~Thienpondt, B.~Desplanques, and
  H.~Na, ``{ECAPA}-{TDNN} embeddings for speaker diarization,'' in
  \emph{INTERSPEECH}.\hskip 1em plus 0.5em minus 0.4em\relax {ISCA}, 2021.

\bibitem{chenglin}
C.~Xu, W.~Rao, J.~Wu, and H.~Li, ``Target speaker verification with selective
  auditory attention for single and multi-talker speech,'' in \emph{{IEEE/ACM}
  Transactions on Audio, Speech, and Language Processing}, vol.~29.\hskip 1em
  plus 0.5em minus 0.4em\relax IEEE/ACM, 2021, pp. 2696--2709.

\bibitem{dehubert}
D.~Ng~\textit{et al.}, ``De’hubert: Disentangling noise in a self-supervised
  model for robust speech recognition,'' in \emph{International Conference on
  Acoustics, Speech and Signal Processing (ICASSP)}.\hskip 1em plus 0.5em minus
  0.4em\relax IEEE, 2023.

\bibitem{chen2023leveraging}
C.~Chen, Y.~Hu, Q.~Zhang, H.~Zou, B.~Zhu, and E.~S. Chng, ``Leveraging
  modality-specific representations for audio-visual speech recognition via
  reinforcement learning,'' in \emph{AAAI Conference on Artificial
  Intelligence}.\hskip 1em plus 0.5em minus 0.4em\relax Association for the
  Advancement of Artificial Intelligence (AAAI), 2023.

\bibitem{cnvmix}
D.~Ng~\textit{et al}., ``Contrastive speech mixup for low-resource keyword
  spotting,'' in \emph{International Conference on Acoustics, Speech and Signal
  Processing (ICASSP)}.\hskip 1em plus 0.5em minus 0.4em\relax IEEE, 2023.

\bibitem{x-vectors}
D.~Snyder, D.~Garcia-Romero, G.~Sell, D.~Povey, and S.~Khudanpur, ``X-vectors:
  Robust dnn embeddings for speaker recognition,'' in \emph{ICASSP}.\hskip 1em
  plus 0.5em minus 0.4em\relax {IEEE}, 2018.

\bibitem{statpool}
M.~Rouvier, P.-M. Bousquet, and J.~Duret, ``Study on the temporal pooling used
  in deep neural networks for speaker verification,'' in \emph{29th European
  Signal Processing Conference (EUSIPCO)}, 2021.

\bibitem{deepsets}
M.~Zaheer, S.~Kottur, S.~Ravanbakhsh, B.~Poczos, R.~R. Salakhutdinov, and A.~J.
  Smola, ``Deep sets,'' in \emph{Advances in Neural Information Processing
  Systems}, vol.~30, 2017.

\bibitem{RawNetv3}
J.~weon Jung, Y.~J. Kim, H.-S. Heo, B.-J. Lee, Y.~Kwon, and J.~S. Chung,
  ``Pushing the limits of raw waveform speaker recognition,'' in
  \emph{INTERSPEECH}.\hskip 1em plus 0.5em minus 0.4em\relax ISCA, 2022.

\bibitem{Kuzmin_2022}
N.~Kuzmin, I.~Fedorov, and A.~Sholokhov, ``Magnitude-aware probabilistic
  speaker embeddings,'' in \emph{The Speaker and Language Recognition Workshop
  (Odyssey 2022)}.\hskip 1em plus 0.5em minus 0.4em\relax {ISCA}, 2022.

\bibitem{MFA-Conformer}
Y.~Zhang, Z.~Lv, H.~Wu, S.~Zhang, P.~Hu, Z.~Wu, H.~yi~Lee, and H.~M. Meng,
  ``{MFA}-conformer: Multi-scale feature aggregation conformer for automatic
  speaker verification,'' in \emph{INTERSPEECH}.\hskip 1em plus 0.5em minus
  0.4em\relax {ISCA}, 2022.

\bibitem{ECAPA-TDNN}
B.~Desplanques, J.~Thienpondt, and K.~Demuynck, ``{ECAPA}-{TDNN}: Emphasized
  channel attention, propagation and aggregation in {TDNN} based speaker
  verification,'' in \emph{INTERSPEECH}.\hskip 1em plus 0.5em minus 0.4em\relax
  {ISCA}, 2020.

\bibitem{attn_stat_pool}
K.~Okabe, T.~Koshinaka, and K.~Shinoda, ``Attentive statistics pooling for deep
  speaker embedding,'' in \emph{INTERSPEECH}.\hskip 1em plus 0.5em minus
  0.4em\relax {ISCA}, 2018.

\bibitem{SEBlock}
J.~Hu, L.~Shen, and G.~Sun, ``Squeeze-and-excitation networks,'' in
  \emph{{IEEE/CVF} Conference on Computer Vision and Pattern
  Recognition}.\hskip 1em plus 0.5em minus 0.4em\relax IEEE/CVF, 2018.

\bibitem{Res2Net}
S.-H. Gao, M.-M. Cheng, K.~Zhao, X.-Y. Zhang, M.-H. Yang, and P.~Torr,
  ``Res2net: A new multi-scale backbone architecture,'' \emph{{IEEE}
  Transactions on Pattern Analysis and Machine Intelligence}, 2021.

\bibitem{TRANSFORMER}
A.~Vaswani~\textit{et al.}, ``Attention is all you need,'' in \emph{Advances in
  Neural Information Processing Systems}, I.~Guyon, U.~V. Luxburg, S.~Bengio,
  H.~Wallach, R.~Fergus, S.~Vishwanathan, and R.~Garnett, Eds., vol.~30.\hskip
  1em plus 0.5em minus 0.4em\relax Curran Associates, Inc., 2017.

\bibitem{subakan2021sepformer}
C.~Subakan, M.~Ravanelli, S.~Cornell, M.~Bronzi, and J.~Zhong, ``Attention is
  all you need in speech separation,'' in \emph{ICASSP 2021-2021 IEEE
  International Conference on Acoustics, Speech and Signal Processing
  (ICASSP)}, 2021, pp. 21--25.

\bibitem{SepformerEnhancement}
D.~de~Oliveira, T.~Peer, and T.~Gerkmann, ``Efficient transformer-based speech
  enhancement using long frames and stft magnitudes,'' in \emph{INTERSPEECH},
  2022.

\bibitem{setAttention}
J.~Lee, Y.~Lee, J.~Kim, A.~Kosiorek, S.~Choi, and Y.~W. Teh, ``Set transformer:
  A framework for attention-based permutation-invariant neural networks,'' in
  \emph{Proceedings of the 36th International Conference on Machine Learning},
  2019, pp. 3744--3753.

\bibitem{Perceiver}
A.~Jaegle, F.~Gimeno, A.~Brock, A.~Zisserman, O.~Vinyals, and J.~Carreira,
  ``Perceiver: General perception with iterative attention,'' in
  \emph{International Conference on Machine Learning}, 2021.

\bibitem{ACANLP}
H.~Wang, C.~Deng, J.~Yan, and D.~Tao, ``Asymmetric cross-guided attention
  network for actor and action video segmentation from natural language
  query,'' in \emph{{IEEE/CVF} International Conference on Computer Vision
  (ICCV)}.\hskip 1em plus 0.5em minus 0.4em\relax IEEE/CVF, 2019.

\bibitem{ACANLP2}
Z.~Ji, J.~Hu, D.~Liu, L.~Y. Wu, and Y.~Zhao, ``Asymmetric cross-scale alignment
  for text-based person search,'' \emph{IEEE Transactions on Multimedia}, pp.
  1--11, 2022.

\bibitem{ACAHNet}
X.~Zhang, S.~Cheng, L.~Wang, and H.~Li, ``Asymmetric cross-attention
  hierarchical network based on cnn and transformer for bitemporal remote
  sensing images change detection,'' \emph{IEEE Transactions on Geoscience and
  Remote Sensing}, vol.~61, pp. 1--15, 2023.

\bibitem{ACAGAAN}
D.~Wu, H.~Li, Y.~Tang, L.~Guo, and H.~Liu, ``Global-guided asymmetric attention
  network for image-text matching,'' \emph{Neurocomputing}, vol. 481, pp.
  77--90, 2022.

\bibitem{PerceiverIO}
A.~Jaegle, S.~Borgeaud, J.-B. Alayrac, C.~Doersch, C.~Ionescu, D.~Ding,
  S.~Koppula, D.~Zoran, A.~Brock, E.~Shelhamer, O.~J. Henaff, M.~Botvinick,
  A.~Zisserman, O.~Vinyals, and J.~Carreira, ``Perceiver {IO}: A general
  architecture for structured inputs \& outputs,'' in \emph{International
  Conference on Learning Representations}, 2022.

\bibitem{PerceiverAR}
C.~Hawthorne~\textit{et al.}, ``General-purpose, long-context autoregressive
  modeling with perceiver{AR},'' in \emph{Proceedings of the 39th International
  Conference on Machine Learning}.\hskip 1em plus 0.5em minus 0.4em\relax PMLR,
  2022.

\bibitem{HiP}
J.~Carreira, S.~Koppula, D.~Zoran, A.~Recasens, C.~Ionescu, O.~Henaff,
  E.~Shelhamer, R.~Arandjelovic, M.~Botvinick, O.~Vinyals, K.~Simonyan,
  A.~Zisserman, and A.~Jaegle, ``Hip: Hierarchical perceiver,'' 2022.

\bibitem{speechbrain}
M.~Ravanelli~\textit{et al}., ``Speechbrain: A general-purpose speech
  toolkit,'' 2021.

\bibitem{wsj0}
J.~S. Garofolo, D.~Graff, D.~Paul, and D.~Pallett, ``{CSR-I (WSJ0)} complete
  {LDC93S6A.}'' in \emph{Electronic Article}, 1993.

\bibitem{spex}
C.~Xu, W.~Rao, E.~S. Chng, and H.~Li, ``{SpEx}: Multi-scale time domain speaker
  extraction network,'' \emph{{IEEE}/{ACM} Transactions on Audio, Speech, and
  Language Processing}, vol.~28, pp. 1370--1384, 2020.

\bibitem{wsj0-2mix}
J.~R. Hershey, Z.~Chen, J.~Le~Roux, and S.~Watanabe, ``Deep clustering:
  Discriminative embeddings for segmentation and separation,'' in \emph{IEEE
  International Conference on Acoustics, Speech and Signal Processing
  (ICASSP)}.\hskip 1em plus 0.5em minus 0.4em\relax IEEE, 2016, pp. 31--35.

\bibitem{adamoptim}
D.~P. Kingma and J.~Ba, ``Adam: {A} method for stochastic optimization,'' in
  \emph{3rd International Conference on Learning Representations, (ICLR)},
  2015.

\bibitem{CLRS}
L.~N. Smith, ``Cyclical learning rates for training neural networks,'' in
  \emph{{IEEE} Winter Conference on Applications of Computer Vision
  (WACV)}.\hskip 1em plus 0.5em minus 0.4em\relax IEEE, 2017, pp. 464--472.

\bibitem{aamsoftmax}
J.~Deng, J.~Guo, J.~Yang, N.~Xue, I.~Cotsia, and S.~P. Zafeiriou, ``{ArcFace}:
  Additive angular margin loss for deep face recognition,'' \emph{{IEEE}
  Transactions on Pattern Analysis and Machine Intelligence}, 2021.

\bibitem{xvecPLDA}
D.~Snyder, P.~Ghahremani, D.~Povey, D.~Garcia-Romero, Y.~Carmiel, and
  S.~Khudanpur, ``Deep neural network-based speaker embeddings for end-to-end
  speaker verification,'' in \emph{IEEE Spoken Language Technology Workshop
  (SLT)}, 2016.

\bibitem{Gu_2021}
W.~Gu, A.~Tandon, Y.-Y. Ahn, and F.~Radicchi, ``Principled approach to the
  selection of the embedding dimension of networks,'' \emph{Nature
  Communications}, vol.~12, no.~1, jun 2021.

\end{thebibliography}

\end{document}